\documentclass[prb,twocolumn,showpacs,amsmath,aps,10pt]{revtex4-1}
\usepackage{graphicx}

\begin{document}

\title{Metal-insulator transition at the 
       $ {\bf LaAlO_3} $/$ {\bf SrTiO_3} $ interface revisited: 
       A hybrid functional study}

\author{F.\ Cossu}
\affiliation{KAUST, PSE Division, Thuwal 23955-6900, Kingdom of Saudi Arabia}
\author{U.\ Schwingenschl\"ogl}
\email[Email: ]{udo.schwingenschlogl@kaust.edu.sa}
\affiliation{KAUST, PSE Division, Thuwal 23955-6900, Kingdom of Saudi Arabia}
\author{V.\ Eyert}
\altaffiliation[Present address: ]{Materials Design sarl, 92120 Montrouge,
                                   France}
\email[Email: ]{veyert@materialsdesign.com}
\affiliation{Experimental Physics VI, 
             Center for Electronic Correlations and Magnetism,
             Institute of Physics, University of Augsburg,
             86135 Augsburg, Germany}

\date{\today}

\begin{abstract}
We investigate the electronic properties of the 
$ {\rm LaAlO_3} $/$ {\rm SrTiO_3} $ interface using density functional 
theory. In contrast to previous studies, which relied on (semi-)local 
functionals and the GGA$ + U $ method, we here use a recently developed 
hybrid functional to determine the electronic structure. This approach  
offers the distinct advantage of accessing both the metallic and insulating
multilayers on a parameter-free equal footing. As compared to calculations 
based on semilocal GGA functionals, our hybrid functional calculations 
lead to a considerably increased band gap for the insulating systems. 
The details of the electronic structure show substantial deviations from
those obtained by GGA calculations. This casts severe doubts on all
previous results based on semilocal functionals. In particular, corrections
using rigid band shifts (''scissors operator'') cannot lead to valid results.
\end{abstract}

\maketitle

\section{Introduction}

One of the most studied heterostructures in recent years is the oxide 
multilayer structure, which is formed from the two bulk insulators 
$ {\rm LaAlO_3} $ (LAO) and $ {\rm SrTiO_3} $ (STO). Unlike its bulk 
constituents, the ground state of the interface has been found to be 
conducting \cite{ohtomo-Nature2004} and magnetic \cite{brinkman-NMat2007}. 
In particular, already early studies reported the formation of an electron 
gas confined to a layer of only a few nanometers at the electron doped 
$ n $-type interface \cite{ohtomo-Nature2006}. The pronounced 
two-dimensional character of this layer has recently been confirmed 
\cite{caviglia-PRL2010}. In general, the nature and spatial confinement 
of two-dimensional electron gases has been the subject of extensive research 
\cite{huijben-NatMat2006,nakagawa-NatMat2006,maurice-PSSA2006,reyren-Science2007,delugas-PRL2011,rubano-PRB2011,safdar-APL2012,berner-PRL2013}.
Interest in the LAO/STO interface stems from the exciting potential for 
applications due to the high mobility of the charge carriers, which 
makes the material a promising building block for field-effect devices
\cite{cheng-Nnano2011,foerg-APL2012}. Following an early suggestion of 
Breitschaft and coworkers \cite{breitschaft-PRB2010} the electron system 
at the $ n $-type interface is now widely regarded as an electron liquid 
due to strong electronic interactions. 

A striking characteristic of the LAO/STO interface is the existence of 
a critical thickness of the LAO overlayer, below which the system is 
insulating. However, the exact number of LAO overlayers necessary 
to drive this thickness-dependent insulator-metal transition is still 
controversially discussed \cite{chen-advmat2010}. Early measurements of 
the sheet carrier density 
indicate that conduction sets in at an LAO thickness of four unit cells 
\cite{thiel-Science2006}. Applying hard x-ray photoelectron spectroscopy 
to samples grown under the same conditions as those of Thiel {\em et al.}, 
Sing {\em et al.}\ find the onset of a finite sheet carrier density already 
at a thickness of two LAO layers \cite{sing-prl2009}. In addition, they 
find a further increase of the sheet carrier density on increasing the 
overlayer thickness. In contrast, Thiel {\em et al.}\ report that the 
sheet carrier density remains essentially constant once the critical 
thickness of four layers has been reached. First principles local density 
approximation (LDA) calculations by Chen {\em et al.}\ confirm the onset 
of a finite conductivity at four LAO layers. Yet, these authors estimate 
that the critical thickness would be rather at five layers or beyond if 
band gap corrections were included \cite{chen-advmat2010}. 

A strong influence of the oxygen content on the sheet carrier density has 
been established by Brinkman {\it et al.}\ \cite{brinkman-NMat2007} as well 
as Kalabukhov {\it et al.}\ \cite{kalabukhov-PRB2007}. Pavlenko {\em et al.}\ 
investigated especially the magnetic ordering at the interface 
\cite{pavlenko-PRB2012a,pavlenko-PRB2012b}. Yet, as has been pointed out 
by Basletic {\it et al.}, the electron gas arising from oxygen vacancies 
is of a pronounced three-dimensional character \cite{basletic-NMat2008}. 

Recent studies \cite{rubano-PRB2011} have confirmed that the formation of 
the two-dimensional electron liquid proceeds in three steps: (1) an orbital 
reconstruction due to broken inversion symmetry and the polarity of the 
interface, (2) polarization triggering a crystal reconstruction by means of 
Ti ion displacement, and (3) creation of a two-dimensional electron liquid 
due to charge from the overlayer. In this context, many theoretical studies 
on the influence of the overlayer thickness on the distortions have been 
performed \cite{pentcheva-PRB2008,pentcheva-PRL2009,udo-EPL2008,udo-EPL2009}. 
The dependence of the structural distortions on the type of interface has 
also been investigated by {\em ab initio} studies \cite{pavlenko-SS2011}. 
The existence of the two-dimensional electron gas has been theoretically 
confirmed \cite{udo-EPL2008,udo-EPL2009,delugas-PRL2011} and was attributed 
to metal-induced gap states \cite{janicka-PRL2009}. 

As a matter of fact, {\em ab initio} calculations with conventional local 
or semilocal exchange-correlation functionals underestimate the optical 
band gap of semiconductors and insulators. Yet, hybrid functionals have 
recently overcome this problem in large part and are known to accurately 
reproduce the experimental band gap of bulk LAO and STO. Hence, they can 
also be expected to better describe the band gap dependence in the LAO/STO 
heterostructure as a function of the overlayer thickness. However, this 
methodology is computationally 
very demanding and, hence, has been rarely applied to large structures  
as the present LAO/STO heterostructure. An exception is the recent 
work by Delugas and coworkers \cite{delugas-PRL2011}. Yet, these 
authors concentrated on periodic LAO/STO heterostructures, where the 
LAO and STO layers are both sandwiched by the respective other material 
and thus the LAO layer is not a true surface layer. 

In the present work we close the gap and employ hybrid functional 
calculations to correctly access the electronic properties of an 
LAO/STO heterostructure with the LAO layer terminated by a vacuum 
region. In doing so, we consider different thicknesses of the LAO 
overlayer in order to determine the critical number of layers, at 
which the insulator-metal transition sets in. While thus going 
beyond most previous studies, our results question the validity of
calculations based on (semi-)local functionals. 

\section{Computational details}

The calculations were performed using the Vienna {\em Ab initio} 
Simulation Package (VASP) \cite{vasp}. The exchange-correlation 
functional was considered at the level of the 
generalized gradient approximation (GGA) \cite{perdew96a}. In addition, 
calculations as based on the recently developed hybrid functonals were 
performed in order to access the electronic properties. Within the 
framework of the generalized Kohn-Sham scheme \cite{seidl96} these 
functionals combine the exchange functional as arising from the LDA 
with the non local Hartree-Fock expression. 
In the present work, the functional proposed by Heyd, Scuseria, and 
Ernzerhof (HSE) was used \cite{hse}. In this approach, the short-range
part of the exchange functional is represented by a (fixed) combination
of GGA- and Hartree-Fock contributions, while the long-range part and
the correlation functional are described by the GGA only.
The single-particle equations were solved using the projector-augmented
wave method \cite{paw,vasppaw} with a plane-wave basis with a
cutoff of 400 eV. The force tolerance was set to 0.05 eV/\AA\ and the 
energy tolerance for the self-consistency loop to 10$^{-5}$ eV.  The 
Brillouin zone was sampled at a grid of $ 4 \times 4 \times 1 $ 
{\bf k}-points.

\begin{figure}
\centering
\includegraphics[width=0.60\columnwidth,clip]{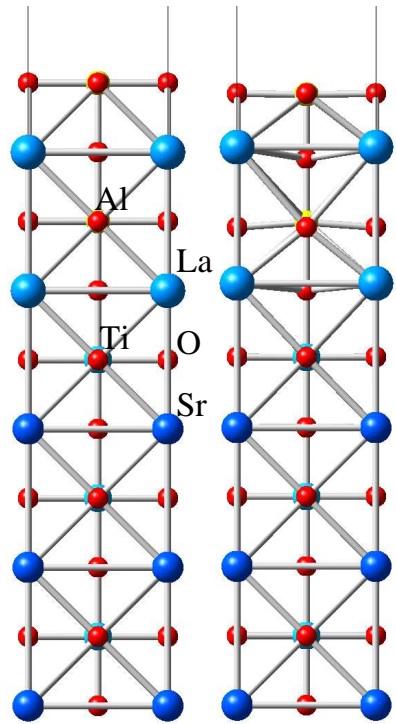}
\caption{Initial idealized (left) and optimized (right) structures with 
         $ 5 \frac{1}{2} $ central layers of $ {\rm SrTiO_3} $ sandwiched 
         between two layers of $ {\rm LaAlO_3} $ from each side.}
\label{fig-sketch}
\end{figure}

\begin{table}[t]
\begin{tabular}{ccccc}
IF 2 & IF 3 & IF 4 & IF 5 & IF 6\\\hline
$+0.025$ & $+0.030$ & $+0.032$ & $+0.036$ & $+0.036$\\
$-0.331$ & $-0.344$ & $-0.337$ & $-0.288$ & $-0.265$\\
$-0.141$ & $-0.116$ & $-0.110$ & $-0.089$ & $-0.074$\\
$-0.193$ & $-0.262$ & $-0.270$ & $-0.230$ & $-0.205$\\
         & $-0.159$ & $-0.133$ & $-0.107$ & $-0.091$\\
         & $-0.187$ & $-0.252$ & $-0.223$ & $-0.196$\\
         &          & $-0.159$ & $-0.114$ & $-0.092$\\
         &          & $-0.183$ & $-0.214$ & $-0.196$\\
         &          &          & $-0.135$ & $-0.094$\\
         &          &          & $-0.146$ & $-0.188$\\
         &          &          &          & $-0.119$\\
         &          &          &          & $-0.131$
\end{tabular}
\caption{\label{tab1} Bucklings (in \AA) of the atomic layers in the LaAlO$_3$
                      region of the systems IF 2 to IF 6, starting from the 
                      surface. Positive/negative signs denote a shift of the 
                      cations with respect to the anions off/to the surface.}
\end{table}

The structural setup consists of a central region of $ {\rm SrTiO_3} $, 
which comprises $ 5 \frac{1}{2} $ unit cells and is terminated by 
$ {\rm TiO_2} $ layers. This region is sandwiched by two to six unit 
cells of $ {\rm LaAlO_3} $ on each side with $ {\rm AlO_2} $ layers at 
the surfaces and, hence, $ {\rm LaO} $ layers at the $ n $-type interfaces.
We will refer to these systems in the following as IF 2 to IF 6. 
The sandwiches are separated by $ \approx 20 $-\AA\ thick vacuum 
layers. As a consequence, the whole slab has inversion symmetry and 
is free of artificial dipoles. The in-plane lattice parameter was 
adopted from an initial full structural optimization of bulk 
$ {\rm SrTiO_3} $ at the GGA level, which resulted in a value of 
3.944 \AA \ and was kept in all calculations. In contrast, the atomic 
positions were all fully relaxed. 

\section{Results}

Starting from ideal structures, which are derived from the geometry of 
bulk $ {\rm SrTiO_3} $ and, hence, consist of perfectly flat atomic 
layers with an equal spacing of 1.972 \AA, we first note an overall 
vertical shrinking of the structures during the structural optimization. 
In addition, we observe distortions of the oxygen octahedra in all three 
parts of the sandwiches. While the displacements of the basal O are 
towards the interface in both the STO and LAO regions, the displacements 
of the apical O are always towards the center of the sandwich, i.e.\ 
away from the interface and towards the interface in the STO and LAO 
regions, respectively. Whereas there is almost no buckling of the layers 
in the central STO region, the buckling is strong in the LAO regions 
with alternating short and long bond lengths between the cations and 
the apical O atoms. In general, the alternation of the bond lengths 
results in an elongation and flattening of all octahdra towards and 
away from the interface, respectively. The buckling in the LAO regions 
is stronger for the $ {\rm LaO} $ layers than for the $ {\rm AlO_2} $ 
layers, with both Al and La atoms being farther away from the interface 
as compared to the basal O atoms of the respective layers. For the 
interface with two LAO layers the initial and optimized structures 
are sketched in Fig.\ \ref{fig-sketch}. In addition, Table \ref{tab1}
summarizes the bucklings encountered in the different systems under
investigation.

On moving away from the interface, the buckling of the $ {\rm LaO} $ and 
$ {\rm AlO_2} $ increases and decreases, respectively. The $ {\rm AlO_2} $ 
surface layer also has an inward relaxation, which is stronger than that 
of the other layers (the inward relaxation, accumulated over all layers,  
determines the shrinking of the slab). Differences arise for the largest 
system under study, which contains six layers of $ {\rm LaAlO_3} $ on either 
side of the sandwich. Here, the $ {\rm TiO_2} $ layers have a weak but 
non-negligible buckling and the surface $ {\rm AlO_2} $ layer has a 
buckling, which is opposite to that of the other $ {\rm AlO_2} $ layers. 
The asymmetry of the octahedra is less pronounced for the LAO regions and 
in the STO region it is opposite to that of the other Ti-centered octahedra.

Since a full structural relaxation is not feasible using hybrid functionals  
due to the huge computational cost, we continue using the optimized structure 
obtained from the GGA calculations. However, the hybrid functional 
calculations still allow one to calculate the forces on the 
atoms. Finite forces obtained from the HSE functional would then indicate 
additional forces due to the improved treatment of the exchange interaction. 
According to our calculations this interaction leads to additional forces 
acting on the Ti and Al atoms, which point towards the interface. In other 
words, these atoms would move towards the center of the octahedra. This 
finding is in line with the results of previous GGA $ + U $ calculations 
\cite{pentcheva-PRB2008}.

\begin{figure}[t]
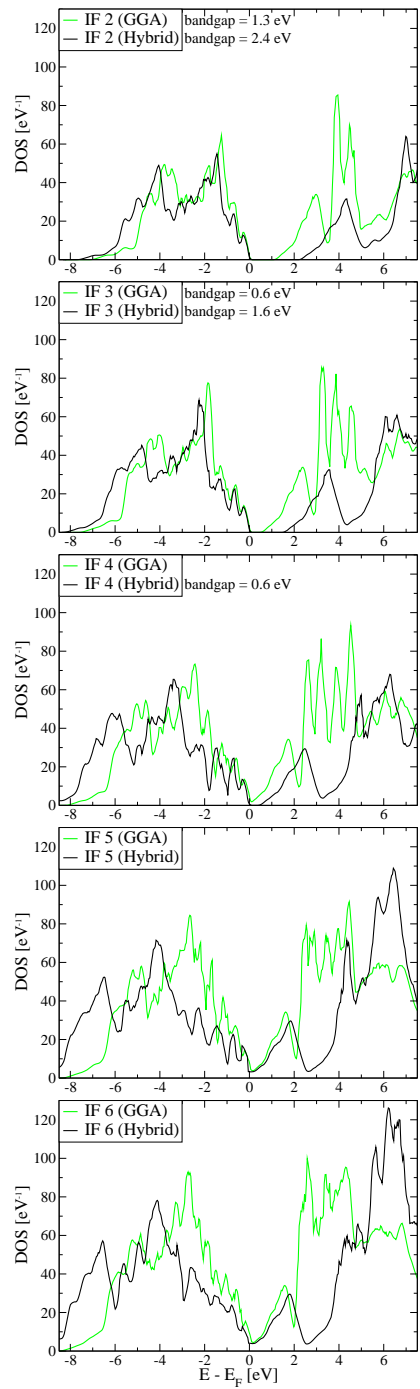

\includegraphics[width=0.3\textwidth,clip]{2.eps}\vspace{-0.25cm}
\includegraphics[width=0.3\textwidth,clip]{3.eps}\vspace{-0.25cm}
\includegraphics[width=0.3\textwidth,clip]{4.eps}\vspace{-0.25cm}
\includegraphics[width=0.3\textwidth,clip]{5.eps}\vspace{-0.25cm}
\includegraphics[width=0.3\textwidth,clip]{6.eps}
\caption{Total DOSs for LAO thicknesses of two to six unit cells obtained in
         the GGA and hybrid functional calculations. The exact gap values 
         are given, for the insulating systems.}
\label{total_DOS-fig}
\end{figure}

According to the total DOS as calculated with the hybrid functional 
and shown in Fig.\ \ref{total_DOS-fig}, 
the insulator-metal 
transition occurs between four and five unit cells of LAO. In contrast, 
the GGA calculation give a metallic state already at four unit cells. 
This is in agreement with existing experimental data whereas the 
optical band gap of 0.6 eV arising from the HSE calculations clearly 
signals insulating behavior. Surprisingly, the situation is not 
unlike that of Ge, where calculations using local or semilocal 
functionals lead to metallic behavior, while hybrid functional 
calculations perfectly reproduce the experimental band gap of 0.66 eV. 
We thus argue that given our calculated results it is worthwhile 
rechecking the experimental data with respect to the exact number of 
layers at the transition.

\begin{figure*}[t]
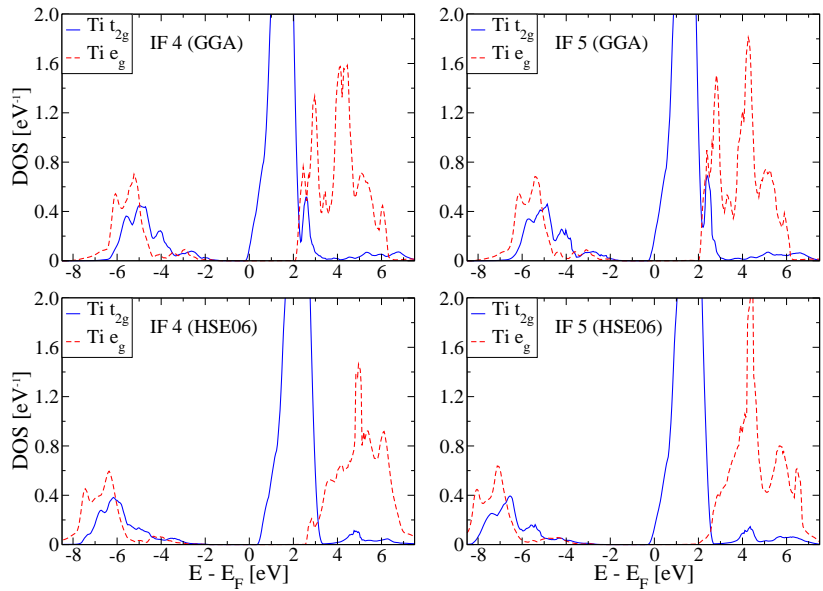

\includegraphics[width=0.3\textwidth,clip]{4-Ti.eps}\hspace{-0.1cm}
\includegraphics[width=0.3\textwidth,clip]{5-Ti.eps} \\\vspace{-0.25cm}
\includegraphics[width=0.3\textwidth,clip]{4-Ti-HF.eps}\hspace{-0.1cm}
\includegraphics[width=0.3\textwidth,clip]{5-Ti-HF.eps}
\caption{Projected $t_{2g}$ and $e_g$ DOSs for the Ti1 atom.}
\label{Ti_DOS-fig}
\end{figure*}

Despite the difference in absolute value the optical band gaps arising 
from the GGA and HSE calculations reveal an almost identical decrease 
of the gap on increasing thickness of the LAO layer. However, the 
differences between both approaches go beyond a rather simple rigid 
band shift. According to Fig.\ \ref{total_DOS-fig} the total densities 
of states show a markable qualitative difference especially for the 
conduction bands. This is explored in more detail in the subsequent 
analysis of the projected densities of states. 

In agreement with the general picture developed in previous work, the 
bottom of the conduction band (which eventually gets pinned to the Fermi 
level) traces back to Ti states, see Fig.~\ref{Ti_DOS-fig}. In particular, 
we note that the $t_{2g}$ states are next to the conduction band minimum 
and therefore are the first to be populated. From the GGA we thus obtain 
quantitative agreement with the findings in Ref.\ \cite{pentcheva-PRB2006}.
The $t_{2g}$ states are responsible for the interface conductivity. On 
the other hand, the $e_g$ states, which are located several eV away, are 
further pushed upwards in energy when the Hartree-Fock exchange is turned 
on. The two-dimensional nature of the electron gas is reflected by the
fact that at the interface the hybridization of the Ti states with the O$_b$ 
(basal O) states is much stronger than with the O$_a$ (apical O) states, for
O atoms bound to the same Ti atom.

\begin{figure*}[t]
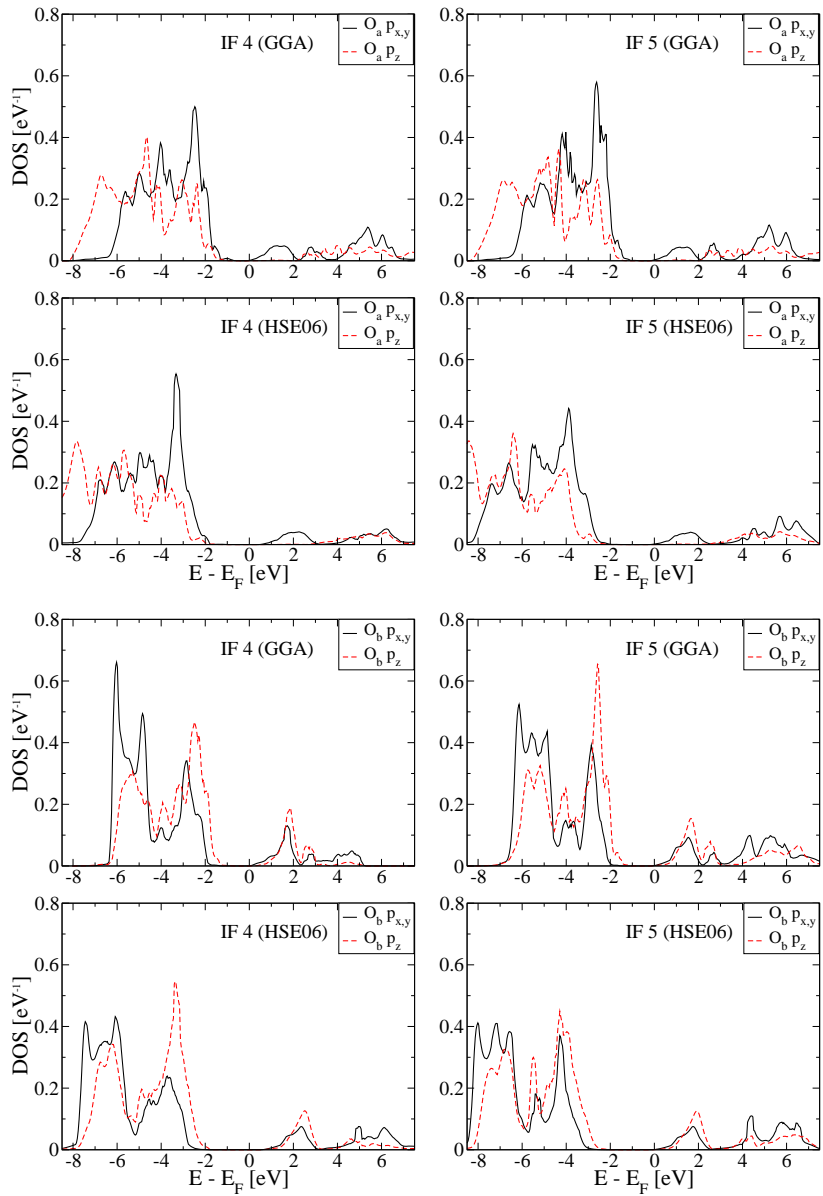

\includegraphics[width=0.3\textwidth,clip]{4-Oa.eps}\hspace{-0.1cm}
\includegraphics[width=0.3\textwidth,clip]{5-Oa.eps} \\\vspace{-0.25cm}
\includegraphics[width=0.3\textwidth,clip]{4-Oa-HF.eps}\hspace{-0.1cm}
\includegraphics[width=0.3\textwidth,clip]{5-Oa-HF.eps} \\\vspace{+0.25cm}
\includegraphics[width=0.3\textwidth,clip]{4-Ob.eps}\hspace{-0.1cm}
\includegraphics[width=0.3\textwidth,clip]{5-Ob.eps} \\\vspace{-0.25cm}
\includegraphics[width=0.3\textwidth,clip]{4-Ob-HF.eps}\hspace{-0.1cm}
\includegraphics[width=0.3\textwidth,clip]{5-Ob-HF.eps}
\caption{Projected DOSs for the apical (O$_a$, between Al and Ti at the 
         interface) and basal (O$_b$, bonded to Al only) O atoms.}
\label{O_DOS-fig}
\end{figure*}

In Fig.\ \ref{O_DOS-fig} 
we present the projected DOS obtained for 
apical (top) and basal (bottom) O atoms, projected on the
in-plane ($p_{x,y}$) and out-of-plane ($p_z$) components of the 
$p$  orbitals. In a perfect octahedral symmetry we would expect 
a similar DOS shape for the in-plane O$_a$ and out-of-plane 
O$_b$ states. Here, we observe at and right above the Fermi 
energy a contribution from the $p_{x,y}$ states of O$_a$. In 
contrast, the $ p_z $ states of the O$_a$ atoms nearly vanish. 
Hence, the $p$ states of the O$_a$ sites right above the Fermi 
energy are of mainly $ t_{2g} $ character as expected. For the 
O$_b$ atoms, we observe similar contributions from the $ p_{x,y} $ 
and $ p_z $ states, which likewise reflect the $ t_{2g} $ symmetry. 
However, their contribution is much larger than that of the O$_a$ 
sites thus clearly confirming the distinct two-dimensional nature 
of the electronic states. This result is the same for GGA and HSE. 

The unoccupied O$_a$ $p_{x,y}$ states at the interface give rise to 
two peaks close to 5 and 6 eV for all metallic systems shown in 
the upper part of Fig.\ \ref{O_DOS-fig}. These peaks are located at 
lower energy in the HSE DOS in the case of the insulating system with 
four LAO unit cells. The out-of-plane O$_b$ $p_z$ orbitals are 
similar to the in-plane O$_a$ $p_{x,y}$ orbitals. For the basal 
O atoms we obtain a similar behavior, but the peaks come from both 
the in-plane and out-of-plane $ p $ orbitals.

\section{Conclusion}

In the present work we have employed semilocal and hybrid functionals to 
compute the equilibrium geometries and electronic ground states of LAO/STO 
interfaces comprised of a central STO region and LAO surface regions with  
varying thicknesses. Our results show that calculations using the GGA and 
HSE functionals predict an insulator-metal transition for thicknesses 
of three to four unit cells and four to five unit cells of LAO, respectively.
While the overall thickness dependence of the electronic structure obtained
from both approaches is similar, the main difference results from the
increased optical band gap. Importantly, the shapes of the (partial) densities
of states show substantial deviations for the GGA and HSE functionals.
Since hybrid functional calculations (even for metallic systems) can be 
expected to describe the shapes with higher accuracy than (semi-)local 
functionals, this fact lays ground for severe doubts on a large number 
of previous results obtained by GGA and GGA$ + U $ calculations. It is 
obvious that band gap corrections using rigid band shifts cannot be 
expected to lead to valid results.

\begin{acknowledgments}
We are particularly grateful for vivid discussions with and support from 
N.\ Pavlenko, T.\ Kopp, and J.\ Mannhart. The calculations were performed
using Material Design's MedeA computational environment. We acknowledge 
Materials Design Inc.\ as well as KAUST IT for providing the computational 
resources for this study. This work was supported by the DFG through TRR 80 
and by the BMBF through project No.\ 03SF0353B.
\end{acknowledgments}

\end{document}